# Rapid and precise distance measurement using balanced cross-correlation of a single frequency-modulated electro-optic comb

*Zijian Wang[1]†, Zhuoren Wan[1]†, Jingwei Luo[1], Yuan Chen[1], Mei Yang[1], Qi Wen[1], Xiuxiu Zhang[1], Zhaoyang Wen[1,2], Shimei Chen[1], Ming Yan[1,2]\*, and Heping Zeng[1,2,3]\**

[1] State Key Laboratory of Precision Spectroscopy, and Hainan Institute, East China Normal University, Shanghai 200062, China

[2] Chongqing Key Laboratory of Precision Optics, Chongqing Institute of East China Normal University, Chongqing 401120, China

[3] Jinan Institute of Quantum Technology, Jinan, Shandong 250101, China

† These authors contribute equally.

**Abstract:**

Ultra-rapid, high-precision distance metrology is critical for both advanced scientific research and practical applications. However, current light detection and ranging technologies struggle to simultaneously achieve high measurement speed, accuracy, and a large non-ambiguity range. Here, we present a time-of-flight optical ranging technique based on a repetition-frequency-modulated femtosecond electro-optic comb and balanced nonlinear cross-correlation detection. In this approach, a target distance is determined as an integer multiple of the comb repetition period. By rapidly sweeping the comb repetition frequency, we achieve absolute distance measurements within 500 ns and real-time displacement tracking at single-pulse resolution (corresponding to a refresh rate of 172 MHz). Furthermore, our system attains an ultimate ranging precision of 5 nm (with 0.3 s integration time). Our method uniquely integrates nanometer-scale precision, megahertz-level refresh rates, and a theoretically unlimited ambiguity range within a single platform, while also supporting multi-target detection. These advances pave the way for high-speed, high-precision ranging systems in emerging applications such as structural health monitoring, industrial manufacturing, and satellite formation flying.

# 1. Introduction

Distance metrology is essential across diverse domains, from structural monitoring of large-scale scientific facilities [1, 2], to high-precision relative positioning in satellite formation flying [3]. Many of these applications require metrological systems capable of simultaneously achieving high precision, large ambiguity ranges, and fast refresh rates [4-6]. Particularly for aerospace applications [7], the additional requirements of real-time velocity determination and multi-target tracking pose significant challenges. Despite their widespread use, conventional light detection and ranging (LiDAR) technologies—whether based on time-of-flight (TOF) [8-10] or coherent detection principles [11-13], cannot simultaneously satisfy all these stringent specifications.

The emergence of optical combs, a coherent light source composed of evenly spaced narrow frequency lines [14], has opened up new opportunities for precision metrology. Notably, dual-comb interferometry [15-18], without moving parts, enables rapid distance measurements via heterodyne detection of two combs with slightly different repetition frequencies [19-24]. This technique measures both the time delay and the phase shift of a detection comb to enhance ranging precision at long distances. Recent progress includes absolute distance measurement over a 113 km open path with a precision of 82 nm achieved within 21 s [25]. However, this technique requires processing a full interferogram to extract distance, limiting real-time performance and imposing high computational loads. Furthermore, its reliance on two tightly phase-locked coherent combs constrains practical applications [25-27].

Balanced cross-correlation (BCC) is a technique that enables ultrafast optical pulse synchronization with attosecond-level timing precision [28]. When combined with a femtosecond (fs) laser comb, it provides an alternative ranging mechanism [29], in which case a target distance is determined by a multiple of the comb repetition period. Compared to other techniques, BCC ranging offers nanometer precision (e.g., 8.7 nm at 10 ms [29]) with minimal post-processing, while maintaining accuracy over long distances without ambiguity or coherence limits. However, the requirement to phase-lock and frequency-count the comb's repetition frequency ($f_r$) imposes both refresh rate limitations and system complexity.

In this work, we present an ultra-rapid, high-precision BCC ranging technique based on a $f_r$-swept fs electro-optic (EO) comb. Our approach measures a target distance by identifying the repetition frequencies that nullify the BCC signal. Unlike mode-locked lasers, our EO comb's $f_r$ is instantaneously set by an RF synthesizer, eliminating the need for phase-locking and frequency counting. This enables a refresh rate of up to 2 MHz for absolute distance

measurements—limited only by the RF synthesizer—and ~170 MHz (i.e., the comb's $f_r$) for tracking relative displacements. Such a high refresh rate facilitates velocity measurements of fast-moving objects. More importantly, the system achieves an ultimate ranging precision of ~5 nm after 0.3 s of integration, maintaining exceptional performance even at extended distances (~14.1 km in fiber) with 60 nm precision within 1 s. Additionally, we demonstrate multi-target distance detection. By integrating these capabilities, we establish a powerful platform for high-speed metrology applications.

## 2. Methods

### 2.1. Basic concept

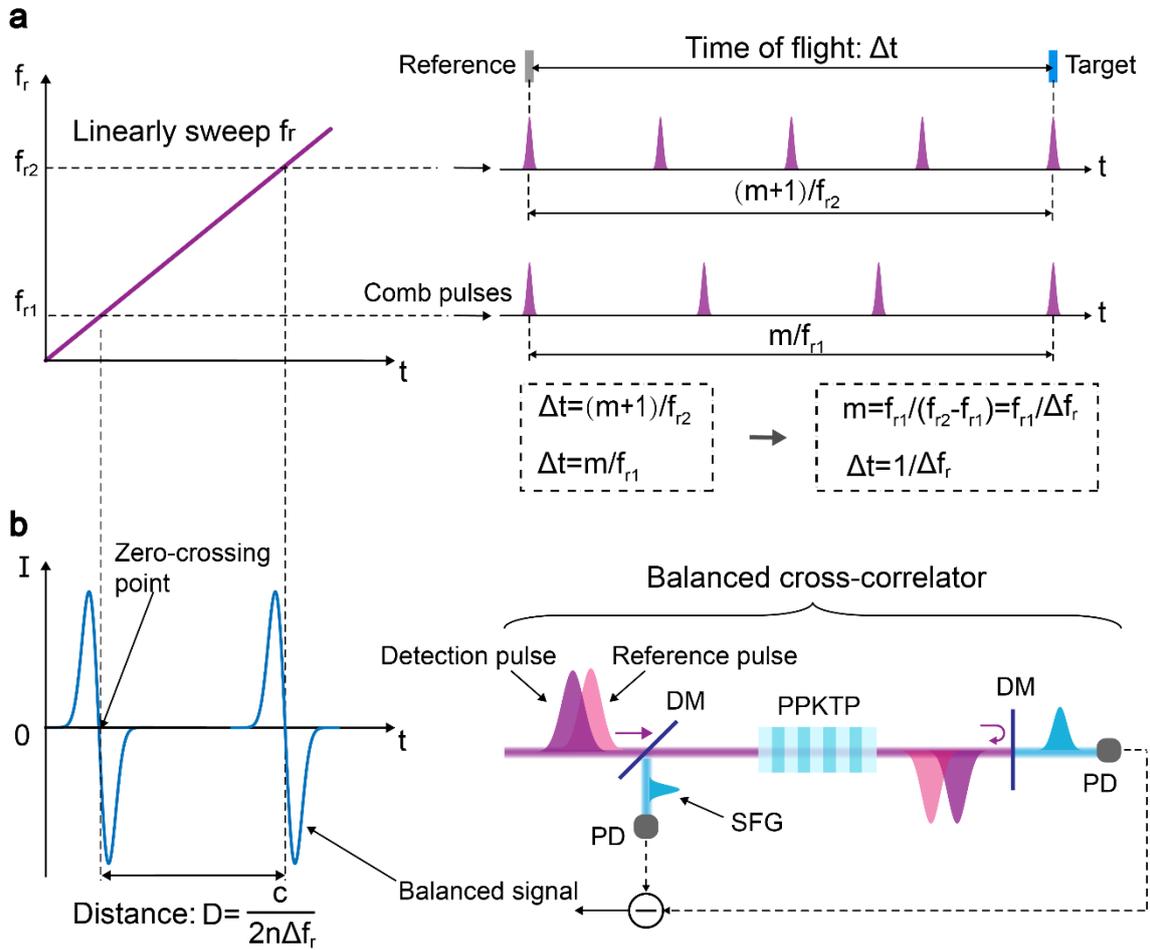

**Figure 1: Working principle. a,** Distance measurement scheme using a repetition-frequency-swept electro-optic (EO) comb. The target distance is encoded in the pulse time-of-flight ($\Delta t$) and extracted via repetition-frequency ($f_r$) tuning. **b,** Detailed balanced cross-correlator design. The balanced signal nullifies when $f_r$ satisfies the condition $\Delta t = m/f_r$, where the integer m is determined stroboscopically. The target distance D is then calculated as $D = c/(2n\Delta f_r)$, where $c$ is the speed of light in vaccum, $n$ is the refractive index of the medium, and $\Delta f_r$ is the repetition

frequency difference. Abbreviations: DM, dichroic mirror; PPKTP, periodically poled KTiOPO$_4$; PD, photodetector.

In our approach, a fs EO comb acts as a time-domain Vernier caliper (**Figure 1(a)**), precisely measuring the travel time ($\Delta t$) between a reference and a target using a balanced cross-correlator (**Figure 1(b)**). Specifically, $\Delta t$ equals an integer multiple of the comb's repetition period ($\Delta t = m/f_r$), occurring when the reflected detection pulse overlaps with the $m$-th reference pulse, nullifying the BCC signal. By linearly sweeping $f_r$ (set by an RF synthesizer) and simultaneously recording the BCC signal with a high-speed digitizer, we identify two adjacent zero-crossing points at $f_{r1}$ and $f_{r2}$. This yields:

$$\Delta t = m/f_{r1} = (m+1)/f_{r2},$$

and thus $\Delta t = 1/\Delta f_r$, where $\Delta f_r = |f_{r2} - f_{r1}|$. The target distance is then:

$$D = c \cdot \Delta t / 2n = c/(2n \cdot \Delta f_r),$$

where $c$ is the speed of light in a vacuum, and $n$ is the reflective index of the air.

Notably, our approach is conceptually analogous to frequency-modulated continuous-wave (FMCW) LiDAR [30-32], if one equates the comb's $f_r$ to the chirped laser frequency ($f_{cw}$) in FMCW. In FMCW LiDAR, the distance is extracted by measuring the beat frequency ($\Delta f_{cw}$) between the chirped laser and its echo via optical heterodyne detection. In contrast, our method employs an $f_r$-swept EO comb and determines $\Delta f_r$ through BCC detection which directly converts $f_r$ into intensity, enabling instantaneous measurements (after calibration). Our approach offers several key advantages: (1) ambiguity-free ranging (theoretically), (2) nanometer-scale precision, and (3) real-time measurement capability with MHz-level refresh rates.

**2.2. Experimental setup**

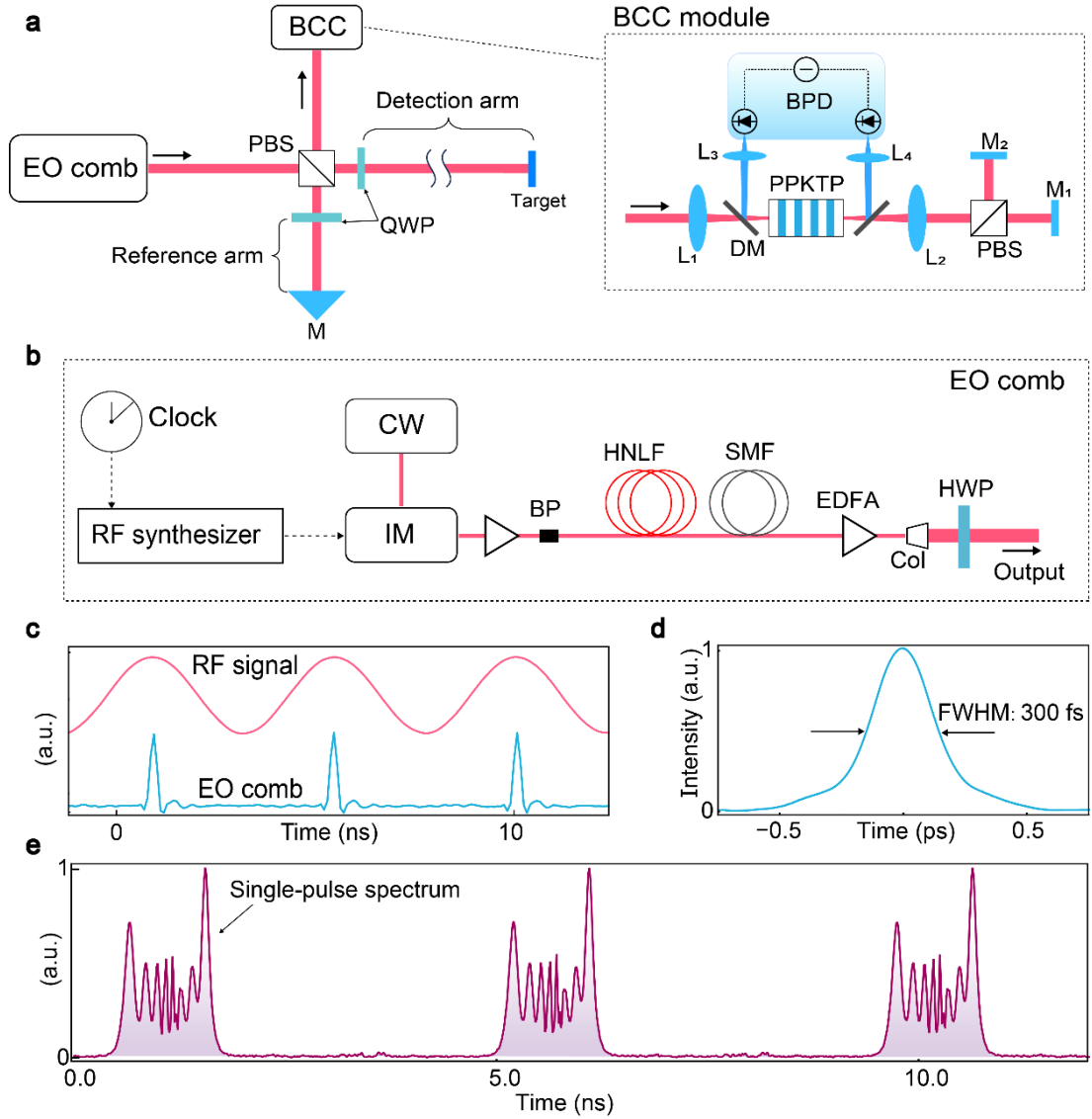

**Figure 2: Experimental setup and characterization. a,** Schematic of the ranging setup. Optical components: BCC, balanced cross-correlator; QWP, quarter-wave plate; PBS, polarizing beamsplitter; M, mirror; L, lens (focal lengths: $L_1$, $L_2$ = 50 mm and $L_3$, $L_4$ = 30 mm); BPD, balanced photodetector. Mirrors $M_1$-$M_2$ provide adjustable delay between cross-polarized components. **b,** Electro-optic (EO) comb generation. IM, intensity modulator; EDFA, erbium-doped fiber amplifier; BP, bandpass filter; HNLF, highly nonlinear fiber (180 m); SMF, single-mode fiber (100 m); Col, collimator; HWP, half-wave plate. **c,** Measured RF drive signal (upper trace) and modulated optical output (lower trace), acquired using a 4 GS/s digital oscilloscope. **d,** Autocorrelation trace measured by a commercial autocorrelator (APE, pulseCheck). **e,** Single-pulse spectra acquired via time-stretch dispersive Fourier transform spectroscopy (0.2 nm resolution, 6.7 nm/ns dispersion slope).

Our experimental setup (**Figure 2(a)**) consists of two primary subsystems: a $f_r$-swept EO comb generator and a BCC detection module. The EO comb (**Figure 2(b)**) begins with a narrow-linewidth (10 kHz) 1550 nm CW laser (NKT Photonics), which is modulated by a 40 GHz bandwidth intensity modulator (Keyang Photonics), comprising both a picosecond pulse driver and an EO modulator. This modulation unit transforms the sinusoidal output (**Figure 2(c)**) from a hydrogen maser-referenced RF synthesizer (frequency stability $2\times10^{-13}$ at 1 s averaging time) into a train of 30 ps optical pulses with 50 μW average power. Subsequent amplification through a custom-built erbium-doped fiber amplifier (EDFA) boosts the pulse power beyond 150 mW before spectral broadening in a 180-meter-long highly nonlinear fiber (YOFC). This nonlinear fiber is engineered with a zero-dispersion wavelength at 1550 nm, a linear dispersion slope of 0.03 ps/(nm²·km), a loss coefficient below 1.5 dB/km, and a nonlinear coefficient exceeding 10 $W^{-1}$ $km^{-1}$. The comb pulses are then temporally compressed to 300 fs pulse duration (full-width at half-maximum, **Figure 2(d)**) using 100 meters of single-mode fiber (SMF). Furthermore, to assess potential pulse-to-pulse spectral variations induced by modulation instability in the long nonlinear fiber [33], we employ single-pulse measurements using a customized time-stretch dispersive Fourier-transform spectrometer [34]. As evidenced in **Figure 2(e)**, our EO comb exhibits excellent spectral stability across pulses (e.g., pulse-to-pulse spectral width variations less than 0.2%).

Comparing to mode-locked laser combs, our EO comb system offers several distinct advantages for BCC ranging applications. First, the $f_r$ can be continuously tuned across a wide range (100-500 MHz), effectively minimizing the dead-zone to $D_0 = c/(2n\cdot\Delta f_{max}) \approx 0.375$ m in air (where $n \approx 1.0003$ and $\Delta f_{max} = 400$ MHz). This represents a three-order-of-magnitude improvement over traditional mode-locked laser systems [29]. This small dead-zone is negligible through proper setting of the reference position. Second, our system supports ultra-rapid $f_r$ scanning at rates up to 2 MHz, enabling real-time, dynamic distance measurements. Third, the comb inherits its exceptional frequency stability directly from the atomic-clock-disciplined RF synthesizer, eliminating the requirement for additional active stabilization systems.

For distance measurement, a polarization beamsplitter (PBS) equally divides the EO comb output into detection and reference arms. In the detection path, the transmitted comb pulses illuminate the target, while the reflected pulses combine with the reference arm pulses at the beamsplitter before being directed into the BCC module.

The BCC module comprises a type II PPKTP crystal and a balanced detector. The combined reference and detection pulses first pass through the crystal, generating an initial sum-frequency generation (SFG) signal at 775 nm. The cross-polarized comb beams are then separated by a PBS, with each subsequently retroreflected by a mirror. The returning beams recombine at the same PBS for a second pass through the crystal, generating a counter-propagating SFG signal. Two dichroic mirrors (DMs) spatially separate the two SFG components before they illuminate the balanced photodetector, which differentially subtracts their photocurrents to generate the BCC output. This signal is acquired by a 12-bit digitizer (1 GHz bandwidth) for processing. The balanced architecture inherently suppresses common-mode optical and electronic noise, endowing the system with exceptional sensitivity to tiny distance variations.

## 3. Results

### 3.1. Ranging precision and accuracy

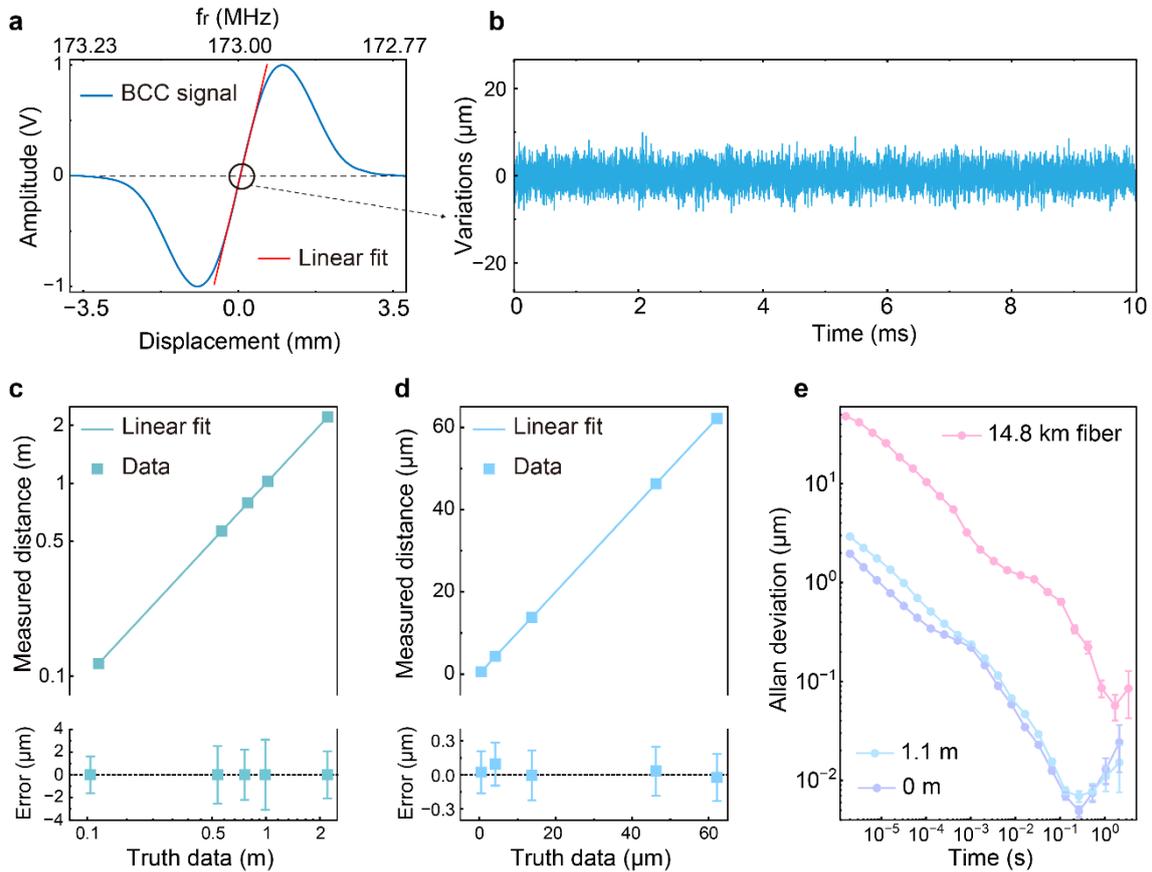

**Figure 3: Evaluation of measurement precision. a,** Representative balanced cross-correlation trace. **b,** Displacement variations derived from amplitude fluctuations. **c,** Comparison of single-shot distance measurements with reference CW laser interferometer data. **d,** Precision enhancement through 100-fold measurement averaging. **e,** Allan deviation analysis showing stability versus integration time.

**Figure 3a** displays a recorded BCC signal generated by linearly sweeping the comb's $f_r$, equivalent to scanning the reference arm. A linear fit to the central region of the BCC signal yields a zero-crossing point at 173 MHz. Further sweeping of $f_r$ reveals a second zero-crossing point at 182.49 MHz, corresponding to a range difference of 15.8 m between the reference and detection arms. This comprises a 9.4-m optical fiber and a free-space path of ~2 m. The slope of the linear fit also provides an amplitude-to-displacement conversion factor of 0.64 mm/V, which we employ to assess ranging uncertainty under fixed $f_r$ conditions. As shown in **Figure 3b**, the resulting 2σ standard deviation (SD) of the ranging measurements is ±3.5 μm.

We measure the displacement of a mirror mounted on a translation stage, comparing the BCC-derived results against truth data from a CW laser interferometer. Statistical analysis of 40 individual measurements per displacement (without averaging) reveals errors within ±3 μm (**Figure 3c**). With 100-fold averaging, the ranging uncertainty reduces to ±300 nm, and the accuracy reaches the sub-100 nm level (**Figure 3d**).

We further characterize the system's stability through Allan deviation measurements at varying target distances, with the comb $f_r$ fixed at a zero-crossing point. Using the amplitude-to-displacement conversion factor, we quantify displacement deviations across different ranges. As shown in **Figure 3e**, the system achieves a ranging precision of 1 μm at 2 μs integration time, improving to ~5 nm for 0.3 s averaging at a 0 m free-space distance. These results reveal the fundamental noise floor of our system, dominated by intrinsic optical and electronic noise sources.

To evaluate the long-distance measurement capability of our system, we employ a 14.1 km fiber spool in a temperature-controlled, vibration-isolated environment to minimize environmental perturbations. By pre-chirping the comb pulses to compensate for fiber dispersion, we achieve a ranging precision of 20 μm at 2 μs integration time, improving to 60 nm with 1 s (Fig. 3e). These results demonstrate the method's potential for remote distance measurement [35] and for calibration of long-haul fiber networks for time-frequency transfer applications [36, 37].

### 3.2. High-speed measurements

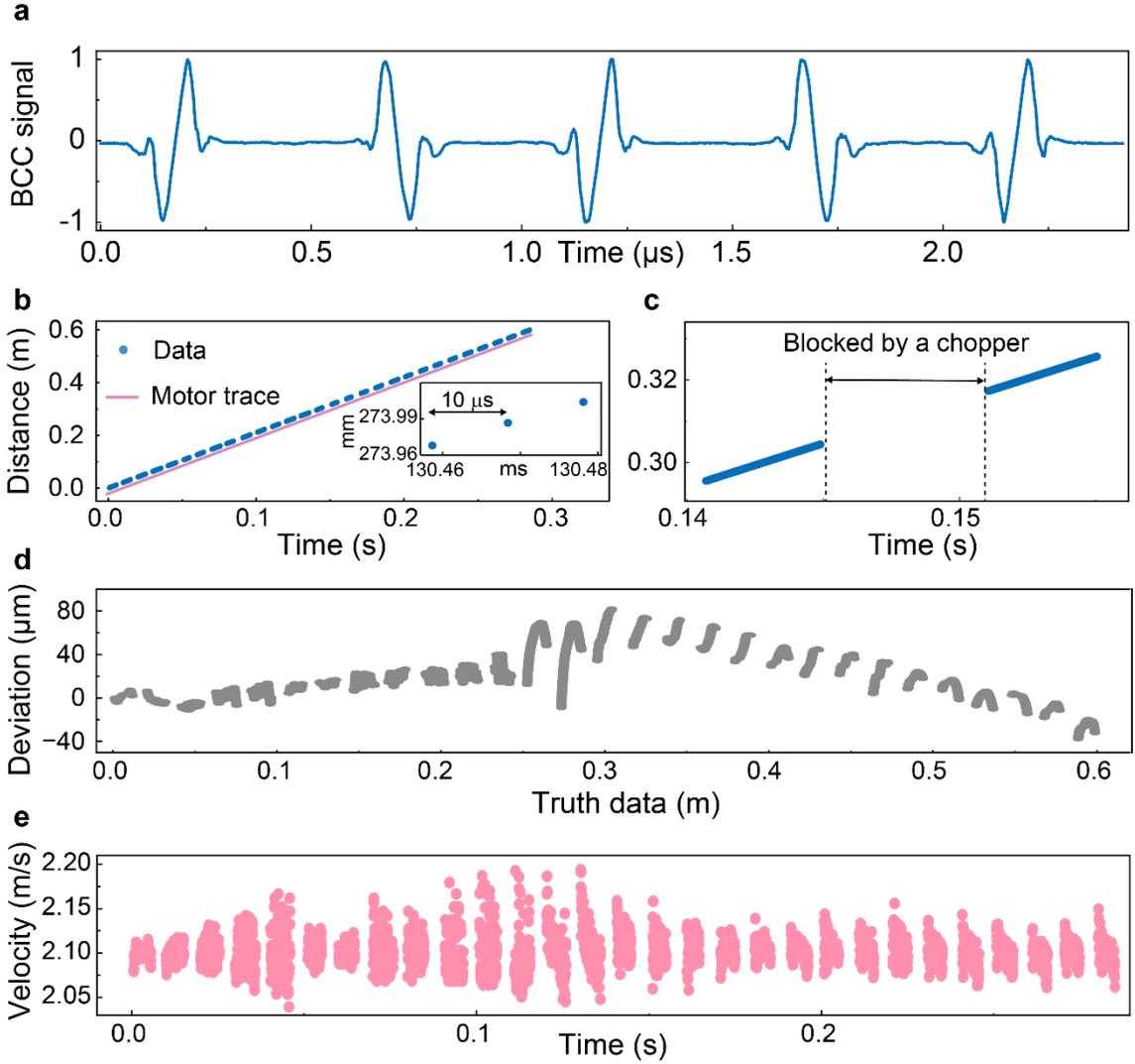

**Figure 4: High-speed displacement and velocity measurements. a,** Balanced cross-correlation signals acquired at a modulation frequency ($f_m$) of 1 MHz. **b,** Displacement measurements of a target moving at 2.1 m/s, measured through a 100 Hz optical chopper (inset: 10 μs sampling interval at $f_m$=100 kHz). **c,** Magnified view of the trajectory measurements from **b**. **d,** Residuals between measured and truth data. The truth data are obtained from a CW interferometer. **e,** Instantaneous velocity data obtained through numerical differentiation of displacement data.

Next, we demonstrate the real-time distance measurement capability by operating the comb in a frequency-modulated mode (modulation frequency, $f_m$; maximum frequency deviation, $\Delta f_{max}$). **Figure 4a** shows the resulting BCC signals ($f_m$ = 1 MHz and $\Delta f_{max}$ = 10 MHz), enabling a 2 MHz refresh rate with zero-crossing detection within only 0.5 μs. Experimentally, we measure a mirror moving at 2.1 m/s by setting $f_m$ =100 kHz and $\Delta f_{max}$ = 10 MHz. A 100 Hz optical chopper in the detection path intentionally interrupts the signal to test system robustness. The measured displacements (blue dots, **Figure 4b**) closely follow the truth trajectory (pink curve, offset vertically for clarity; 10 μs sampling interval shown in inset). The successful recovery of position data during chopped periods (**Figure 4c**) demonstrates the method's resilience to signal interruption—a critical advantage over conventional CW interferometers.

**Figure 4d** quantifies the deviation between measured and truth trajectories, showing a maximum discrepancy of ~80 μm in a folded optical path of 46.4 m. This corresponds to a relative ranging uncertainty of $1.7 \times 10^{-6}$, primarily limited by mechanical vibration from the moving stage. Nevertheless, the system's high refresh rate enables direct derivation of instantaneous velocity from ranging data derivatives (**Figure 4e**), demonstrating the capability for high-speed measurements of distance and velocity.

**3.3. Multipoint ranging**

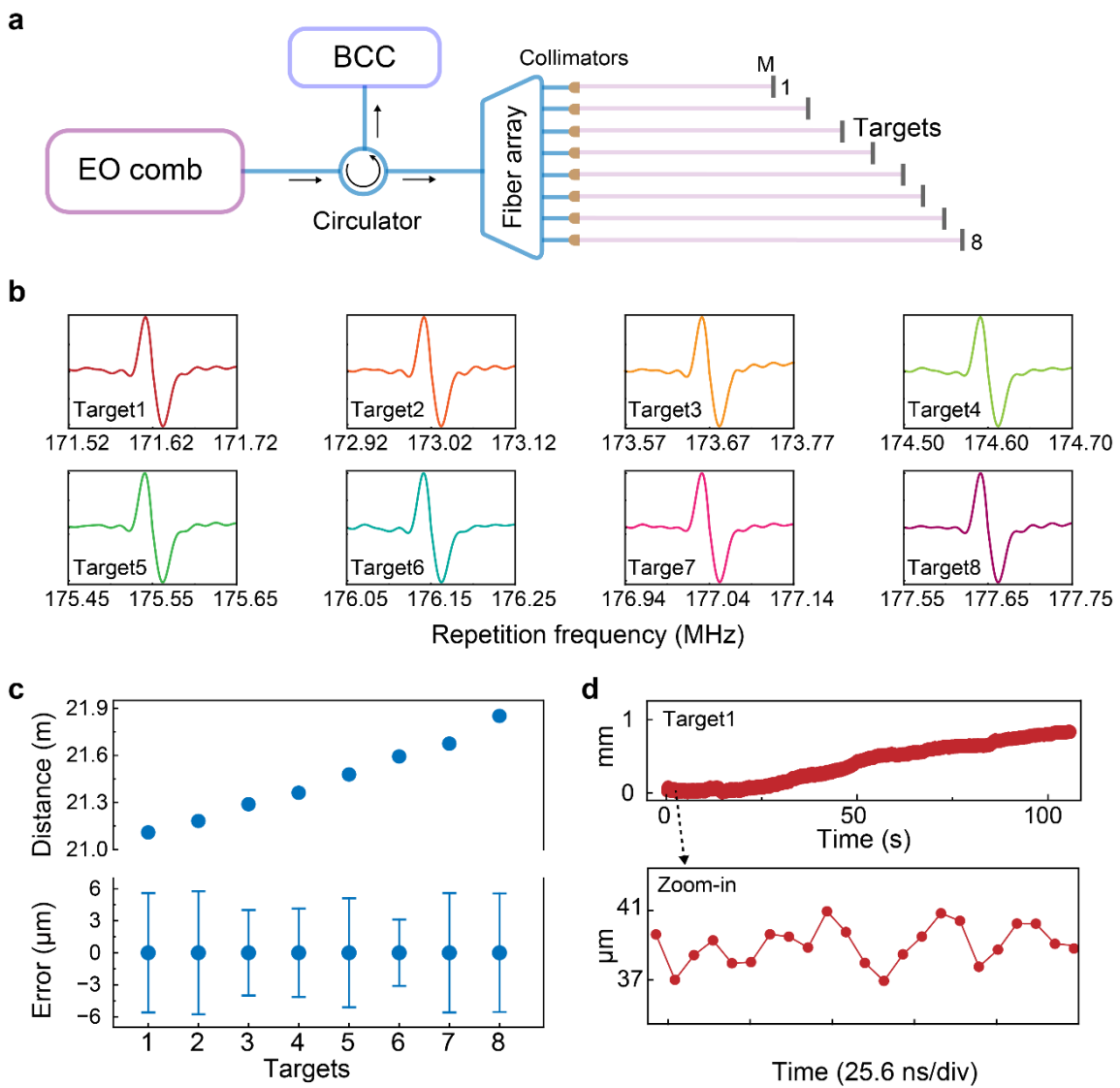

**Figure 5: Multipoint distance measurement demonstration. a** Schematic of the multi-target measurement setup. **b,** Balanced cross-correlation signals acquired from eight spatially separated targets. **c,** Measured absolute distances for all targets relative to the reference arm. **d**, High-speed displacement tracking of Target 1 at the comb repetition rate.

Finally, we demonstrate our method's multipoint measurement capability using a scanner-free multiplexed ranging configuration (**Figure 5a**). The EO comb output is distributed through a 1×8 fiber array, generating eight spatially separated probe beams for multi-target interrogation. Reflected signals are collected through the same array and routed to the BCC module via a fiber circulator.

**Figure 5b** displays the BCC signals acquired from all eight measurement channels as the comb $f_r$ is swept from 170 to 180 MHz in 10 μs. Their correspondence to the targets is established through sequential blocking of individual channels. While the comb power division across eight channels reduces the per-channel signal-to-noise ratio (SNR ≈ 80, defined as peak signal to noise floor), this remains sufficient for precise distance determination relative to the common reference arm (**Figure 5c**). The SNR limitation could be addressed by implementing a high-power fiber amplifier prior to beam splitting.

Importantly, compared to dual-comb techniques, our approach enables both initial distance mapping and subsequent high-speed tracking of selected targets. Following an initial scan to identify all targets (Figure 5b), we can selectively monitor individual targets by setting the corresponding $f_r$ and then tracking their movements at a refresh rate up to the pulse repetition rate. As demonstration, we set $f_r$ at 171.62 MHz to track target 1 (a drifting mirror on a powered-off translation stage) by converting BCC amplitude fluctuations to displacement via a calibrated factor (**Figure 5d**). The magnified view shows the sampling interval at ~172 MHz, highlighting the system's unique combination of multiplexing capacity and high-speed tracking capability.

We note that our time-division multiplexing scheme exhibits a longitudinal resolution limit of ~1.7 mm (determined by the BCC signal width), resulting in aliasing when measuring closely spaced targets. Two potential solutions exist: (1) alternative blocking of the closely spaced signals, or (2) implementing a spatially multiplexed architecture using a shared PPKTP crystal with dual detector arrays—though this would increase system complexity and potentially reduce refresh rates.

## 4. Discussion

Our work introduces a novel frequency-modulated ranging technique that uniquely integrates BCC detection and fs EO comb synthesis. Compared to conventional BCC ranging schemes based on mode-locked laser combs, our approach achieves a significantly smaller dead zone and higher measurement speeds, as detailed in Table 1. Additionally, the system exhibits reduced complexity and enhanced environmental robustness. Notably, recent advancements in lithium niobate photonics have enabled on-chip fs EO comb generation [38-40], making our technique a compelling option for compact LiDAR applications.

Furthermore, a comprehensive comparison between our method and existing ranging techniques is provided in Table 2. Our approach uniquely enables ultra-rapid, high-precision distance measurements while overcoming traditional limitations of comb coherence and range ambiguity. However, challenges remain in laser power consumption and photon detection sensitivity. These limitations could potentially be addressed through emerging photonic technologies, including low-noise high-power fiber amplifiers [41] and ultra-sensitive avalanche photodiodes [42]. Particularly promising is the development of superconducting nanowire detectors [43], which offer single-photon sensitivity. Such advancements could enable ultra-long-range detection and low-light applications like satellite ranging, significantly expanding the potential applications of our system.

## 5. Conclusion

In conclusion, we introduce a cross-correlation-based ranging technique enabled by a frequency-modulated fs EO comb. By integrating high-speed BCC detection with precise frequency modulation, our system achieves a refresh rate of up to 2 MHz and nm-level precision. Our approach provides a theoretically unlimited ambiguity range, eliminating traditional constraints of comb coherence and ambiguity. With these features, our method is promising for diverse metrological and sensing applications.

Table 1 Comparison of $f_r$-modulated and conventional BCC ranging

| Parameters | $f_r$-modulated BCC ranging (This work) | Conventional BCC ranging (ref. 29) |
|---|---|---|
| Sources | Femtosecond EO comb | Mode-locked laser comb |
| Methods | Linear sweep of $f_r$ and simultaneous BCC signal recording | Phase-lock $f_r$ at BCC zero-crossing point |
| $f_r$ determination | Directly set by RF synthesizer | Counted by digital frequency counter |
| $f_r$ tuning range | 100-500 MHz | ±200 kHz |
| Dead zone | 0.375 m | 187.5 m |
| Refresh rate | 2 MHz (0.5 μs) | 4 kHz (0.25 ms) |
| Precision | 5 nm @0.3s | 1.1 nm @1s |

Table 2 Comparison of different ranging techniques

| Parameters | $f_r$-modulated BCC (This work) | FMCW | Pulsed TOF | Dual-comb ranging |
|---|---|---|---|---|
| Laser source | Single EO comb | Single CW laser | Single pulsed laser | Two optical combs |
| Detection | Sum-frequency generation | Heterodyne | Direct detection | Multi-heterodyne |
| Range ambiguity | Theoretically unrestricted | $c \cdot T_{mod}/(2n)$ | $c/(2n \cdot f_r)$ | $c/(2n \cdot \Delta f_r)$ |
| Minimum data refresh time | Pulse repetition period ($1/f_r$) | Modulation period ($T_{mod}$) | $1/f_r$ | Duration of a full interferogram |
| Ranging precision | nm level | mm level | mm level | nm level |
| Coherence requirement | Low | Medium | Low | High |
| Power requirement | High | Low | Low | Medium |
| Full data acquisition | Not needed after calibration | Required | Not needed | Required |